\documentclass[%
 aps,
 jmp,%
 amsmath,amssymb,
 reprint,%
]{revtex4-1}
\usepackage{amsmath,amssymb}
\usepackage{braket}
\usepackage{graphics}
\usepackage{graphicx}
\usepackage{dcolumn}
\usepackage{bm}
\begin{document}
\preprint{AIP/123-QED}
\title[Theoretical study on the slit experiments in the Rashba electron systems]{Theoretical study on the slit experiments in the Rashba electron systems\footnote{Error!}}
\author{Kotaro Shimizu}
\affiliation{Department of Physics, Waseda University, Okubo, Shinjuku-ku, Tokyo 169-8555, Japan}
\author{Masahito Mochizuki}%
\affiliation{Department of Applied Physics, Waseda University, Okubo, Shinjuku-ku, Tokyo 169-8555, Japan}
\date{\today}

\begin{abstract}
We develop a theory of designing slit experiments in two-dimensional electron systems with the Rashba spin-orbit interaction. By simulating the spatiotemporal dynamics of electrons passing through a single slit or a double slit, we find that the interference fringes of the electron probability density attain specific spin orientations via the precession of spins around effective magnetic fields mediated by the Rashba spin-orbit interaction whose directions are determined by the propagation path. The spin orientations of the fringes can be controlled by tuning the Rashba spin-orbit parameter, which can be achieved by applying an electric gate voltage. This phenomenon can be exploited to implement electrically tunable transmission of spin information and to generate spin-polarized currents.
\end{abstract}
\maketitle

Spintronics has attracted a great deal of research interest recently because of its high potential for application to next-generation storage and logic devices. To control the charge and spin of electrons in spintronics devices, the Rashba spin-orbit interaction (RSOI) of relativistic origin plays an extremely important role. This interaction appears in two-dimensional electron systems with broken spatial inversion symmetry~\cite{Rashba60,Bychkov84,Manchon15,Kohda08,Jungwirth12}, and its magnitude can be tuned electrically by application of an electric gate voltage through modulating the asymmetry~\cite{Nitta97,Engels97,Schultz96}.

The Hamiltonian of the RSOI is given by,
$\mathcal{H}_{\rm R}=\hat{\bm{\sigma}}\cdot(\bm{k}\times\bm{\alpha})$,
where $\bm{\alpha}$ is the Rashba vector. This interaction induces an effective magnetic field $\bm H_{\rm RSOI}$ that acts on conduction electrons. The $\bm H_{\rm RSOI}$ field that an electron feels is oriented perpendicular to the momentum (or to the propagation direction) of the electron, which enables us more complex manipulation of their spins than usual magnetic fields. To date, a variety of intriguing physical phenomena and useful device functions associated with RSOI have been reported or proposed~\cite{Manchon15}, such as the spin Hall effect (charge-current to spin-current conversion)~\cite{Sinova04,Kato04}, the spin-galvanic effect~\cite{Ganichev02}, the Edelstein effect~\cite{Edelstein90}, the spin-polarized field-effect transistor~\cite{Datta90,Nitta99}, the spin-filter effect~\cite{Ohe05,Kohda12}, the spin-dependent plasmon excitations (magnetoplasmons)~\cite{Kushwaha06}, and the spin polarizer with circularly polarized light~\cite{Mochizuki18}. Research for novel spin-related transport phenomena associated with the RSOI is an issue of vital importance in the field of spintronics.

In this Letter, we theoretically propose that efficient manipulation of spins can be achieved by exploiting the interference effect of conduction electrons passing through a single slit or a double slit in a Rashba electron system. It is known that the double-slit experiment is one of the epoch-making experiments in physics, which demonstrated de Broglie's wave-particle duality and provided a firm basis to the modern quantum mechanics. In this experiment, however, only the charge degrees of freedom of electrons were concerned. On the contrary, we propose that the interference fringes of electrons in a RSOI system become spin polarized, which enables the spin separation in space. This effect provides a simple means to use electron spins as information carriers and to generate spin-polarized currents~\cite{SpinCurrent12}. 

We simulate the time evolution of a Gaussian wave packet passing through a single slit or a double slit by numerically solving the time-dependent Schr\"{o}dinger equation for a two-dimensional RSOI system. The time-dependent Schr\"{o}dinger equation reads,
\begin{eqnarray}
i\hbar\frac{\partial}{\partial t}\vec{\psi}(\bm r,t)=
\left[\frac{\hat{\bm p}^2}{2m}+\frac{1}{\hbar}\hat{\bm \sigma}\cdot\left(
\hat{\bm p}\times\bm{\alpha}\right)+V\right]
\vec{\psi}(\bm r,t),
\label{eq:Schrodingereq}
\end{eqnarray}
where $\vec{\psi}(\bm r,t)=(\psi_{\uparrow}(\bm r,t),\psi_{\downarrow}(\bm r,t))^{\rm t}$ is the electron wave function, $\hat{\bm \sigma}=(\hat{\sigma}_x,\hat{\sigma}_y,\hat{\sigma}_z)^{\rm t}$ are the Pauli matrices, and $\hat{\bm p}(=\hbar \bm k)$ is the electron momentum. The first and the second terms on the right-hand side give the kinetic energy and the RSOI. We consider a system with RSOI present only in the right-hand side of the slit. The last term $V$ is the rigid-body potential which describes the slit wall. The potential $V$ is set to be 2500 times larger than a kinetic energy of the incident wave packet, while the numerical results are insensitive to the detailed value of $V$ as long as it is large enough. Here $m$ is the mass of free electron, $\hbar$ is the Dirac constant, $\bm{\alpha}$ is the Rashba vector. This kind of system can be realized using an asymmetric heterostructure of semiconductors. We consider a heterostructure in which the spatial inversion symmetry is uniformly broken and thus the Rashba vector is given in the form $\bm{\alpha}=\alpha \bm{e}_{z}$ with $\bm{e}_{z}$ being the unit vector along $z$. We assume a typical value for the Rashba parameter $\alpha$($=\alpha_0=5.06$ meV$\cdot$nm) but also examine effects of its variation. The initial form of the wave packet $\vec{\psi}(\bm{r},t=0)$ is given by,
\begin{eqnarray}
\vec{\psi}(\bm{r},0)\!=\!
\frac{1}{\sqrt{2\pi\delta^2}}\exp\!\!
\left[-\frac{(x-x_0)^2+y^2}{4\delta^2}+ik_{\rm in} x\right] \vec{\chi}(0),
\label{eq:initialstat}
\end{eqnarray}
where $\delta(=$2.7 nm$)$ is the width of the wave packet, and $\bm k_{\rm in}$=$(k_{\rm in},0,0)$ is the incident wave number. We usually consider the case with $k_{\rm in}$=$k_0$(=$0.4\pi/a$) where $a$(=0.6 nm) is the lattice constant but also study effects of its variation. The distance between the initial position of the wave packet and the slit wall is $|x_0|$=60 nm. The quantization axis of the spin is chosen to be the $z$ axis. A recent theoretical study predicted that an electron injected to a point contact with RSOI becomes spin $y$-polarized due to the Landau-Zener tunneling effect~\cite{Iwasaki17}. Thus, we consider the $+y$-polarized spin for the initial spin state $\vec{\chi}(t=0)$ as
$\vec{\chi}(0)=(1/\sqrt{2}, i/\sqrt{2})^{\rm t}$. Note that even if the initial spin polarization deviates from the $y$ direction, the spin-precession frequency will not change because it is governed only by the RSOI-mediated effective magnetic field as argued later, whereas the fringe pattern will vary.

We numerically analyze the time evolution in Eq.~(\ref{eq:Schrodingereq}) by using the fourth-order Runge-Kutta method. For the simulations, we use a system of 600 nm $\times$ 600 nm with open boundary conditions. This system is divided into identical square cells of $a \times a$, and the Laplacian in the equation is treated within the finite-difference method. Using the time evolution of the wave function obtained by the numerical simulations, we calculate spatiotemporal profiles of the dynamical charge and spin densities, $D_{\rm c}=|\vec{\psi}(\bm r,t)|^2$ and $D_{\rm s}=\vec{\psi}(\bm r,t)^{\dag}\hat{\sigma_{z}}\vec{\psi}(\bm r,t)$.
We also study the spin orientation at each interference fringe by calculating the normalized spin-polarized vectors,
\begin{eqnarray}
\bm{n}(\bm r,t)=\frac{\vec{\psi}(\bm r,t)^{\dag}\hat{\bm{\sigma}}\vec{\psi}(\bm r,t)}{|\vec{\psi}(\bm r,t)^{\dag}\hat{\bm{\sigma}}\vec{\psi}(\bm r,t)|}. 
\label{eq:polarvec}
\end{eqnarray}

\begin{figure}[tb]
\includegraphics[scale=0.5]{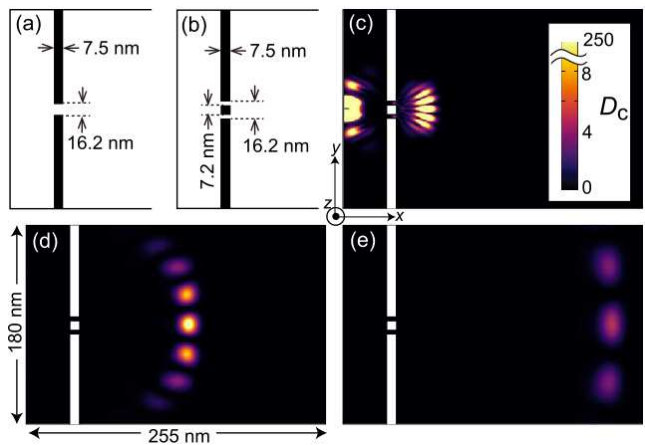}
\caption{(a),(b) Schematic illustrations of (a) the single-slit system and (b) the double-slit system considered herein. (c)-(e) Snapshots of the typical spatiotemporal evolution of the charge-density distribution $D_{\rm c}$ obtained from the simulation at (c)~$t$=0.488 ps, (d)~$t$=0.859 ps, and (e)~$t$=1.33 ps.}
\label{Fig1}
\end{figure}
Figures~\ref{Fig1}(a) and \ref{Fig1}(b) show details of the setup considered in this study, i.e., the single-slit system and the double-slit system, respectively. Figures~\ref{Fig1}(c)-\ref{Fig1}(e) show snapshots of the simulated dynamical charge density $D_{\rm c}$ at selected times for the double-slit case. The results show that after passing through the slit, the electron wave packet exhibits clear interference fringes. In fact, these interference fringes appear even in a system without RSOI. The peculiarities of the RSOI system appear in the spin density $D_{\rm s}$.

\begin{figure*}[tb]
\includegraphics[scale=0.5]{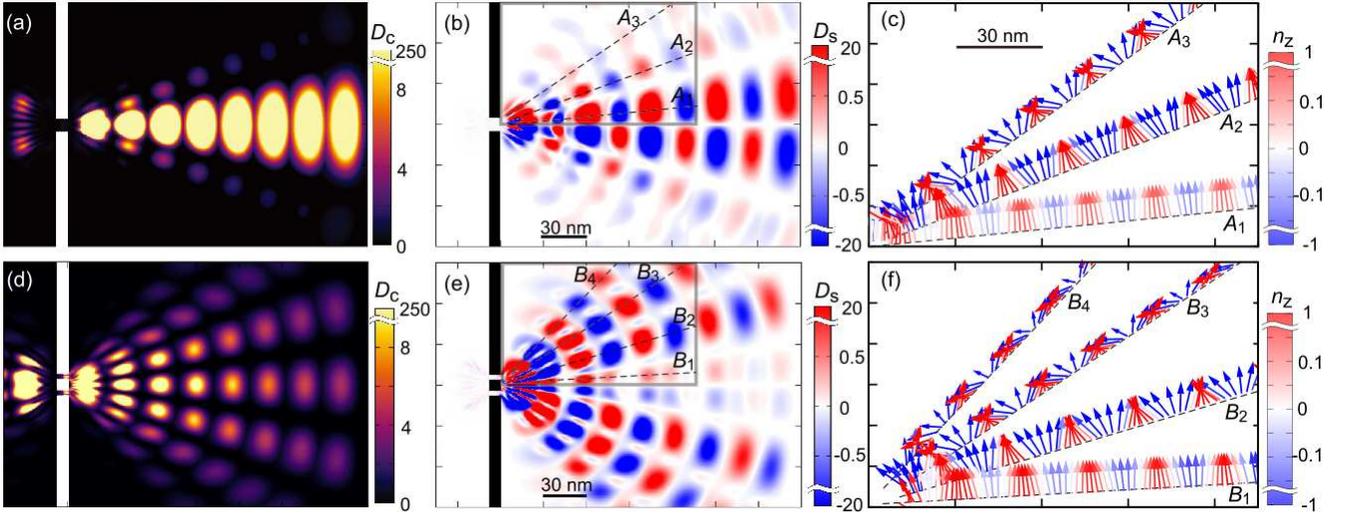}
\caption{(a)-(c) Superimposed images in which all the time-slice snapshots obtained in a simulation are overlaid for spatiotemporal evolutions of (a) the charge density $D_{\rm c}$, (b) the spin density $D_{\rm s}$, and (c) the spin-polarization vectors $\bm n$ on the specific lines indicated by dashed lines in panel (b) for the single-slit experiment with $\alpha=\alpha_0$. The arrows (colors) in panel (c) represent the in-plane (out-of-plane) components of $\bm n$. (d)-(f) Results obtained in a simulation of the double-slit experiment with $\alpha=\alpha_0$.}
\label{Fig2}
\end{figure*}
Figures~\ref{Fig2}(a)-\ref{Fig2}(c) show superimposed images in which all the time-slice snapshots obtained in the simulation are merged for $D_{\rm c}$, $D_{\rm s}$, and $\bm n$, respectively, on specific lines for the single-slit system. Here the simulations were done by using a large system of 600 nm $\times$ 600 nm to avoid influences from the edges, whereas the figures focus on an area of 180 nm $\times$ 255 nm near the slit. We also show simulated results for the double-slit system in Figs.~\ref{Fig2}(d)-\ref{Fig2}(f). From these results, we immediately realize the following three features. First, the spin density $D_{\rm s}$ exactly vanishes in the direction parallel to the initial momentum, whereas the charge density $D_{\rm c}$ is maximized in this direction. Second, both the charge and spin densities $D_{\rm c}$ and $D_{\rm s}$ exhibit interference fringes, and the fringes appear even in the single-slit case because the slit width is sufficiently narrow as compared to the width of the incident Gaussian wave packet. Third, the interference fringes of $D_{\rm s}$ change their signs in an oscillatory manner and with constant intervals. Moreover, the oscillation phases in the upper and lower halves of the system have opposite signs for both the single-slit and the double-slit cases.

\begin{figure}[tb]
\includegraphics[scale=0.4]{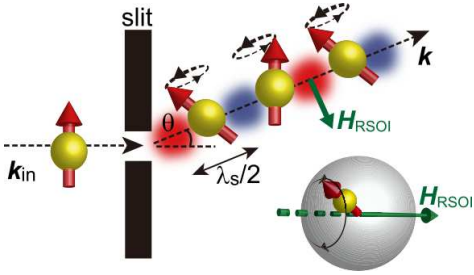}
\caption{Schematic illustration of the behavior of an electron passing through a slit. $\bm{H}_{\rm RSOI}$ denotes the RSOI-induced effective magnetic field. The spin density $D_{\rm s}$ is positive and negative ($D_{\rm s}>0$ and $D_{\rm s}<0$) in the red and blue areas, respectively.}
\label{Fig3}
\end{figure}
The oscillatory behavior of $D_{\rm s}$ can be ascribed to the spin precession around the effective magnetic field induced by the RSOI. Comparing the RSOI term in Eq.~(\ref{eq:Schrodingereq}) with the usual Zeeman term, we find that the RSOI induces an effective magnetic field $\bm H_{\rm RSOI}$ proportional to $\bm{k}\times\bm{\alpha}$. The orientation of the $\bm H_{\rm RSOI}$ field is perpendicularl to the electron propagation direction within the plane. Figure~\ref{Fig3} shows a schematic diagram of the electron spin precession around the $\bm H_{\rm RSOI}$ field. The observed oscillation of $D_{\rm s}$ is caused by the non-adiabatic change in the electron momentum or propagation direction after passing through the slit. In fact, the electrons propagating in the $x$ direction feel the $\bm H_{\rm RSOI}$ field parallel to the original spin direction $\bm y$, and hence spin precession never happens. Conversely, the electrons moving in a direction slanted with respect to the $x$-axis feel the $\bm H_{\rm RSOI}$ field tilted from the $y$ axis. Eventually the spin precession occurs and the dynamical spin $z$-axis components appear.

The space-time dependence of $D_{\rm s}(\bm r,t)$ is approximately described by
$D_{\rm s}(\bm r,t)\propto D_{\rm c}(\bm r,t)\sin\theta\sin(\omega t-\delta)$
where $\theta$ is the angle between electron propagation directions before and after passing through the slit (see Fig.~\ref{Fig3}). The amplitude of the spin precession is determined by the $x$-axis component of the $\bm H_{\rm RSOI}$ field, which is proportional to $\sin\theta$. In the single-slit case, the charge density $D_{\rm c}$ takes large values around the $x$ axis ($\theta \approx 0^\circ$). This large $D_{\rm c}$ causes the observed large oscillation amplitude of $D_{\rm s}$ although $\sin\theta$ is small around $\theta \approx 0^\circ$. Conversely, in the double-slit case, the charge density $D_{\rm c}$ is not so large near the $x$ axis, which results in a very small oscillation amplitude of $D_{\rm s}(\propto \sin\theta)$.

\begin{figure}[tb]
\includegraphics[scale=0.5]{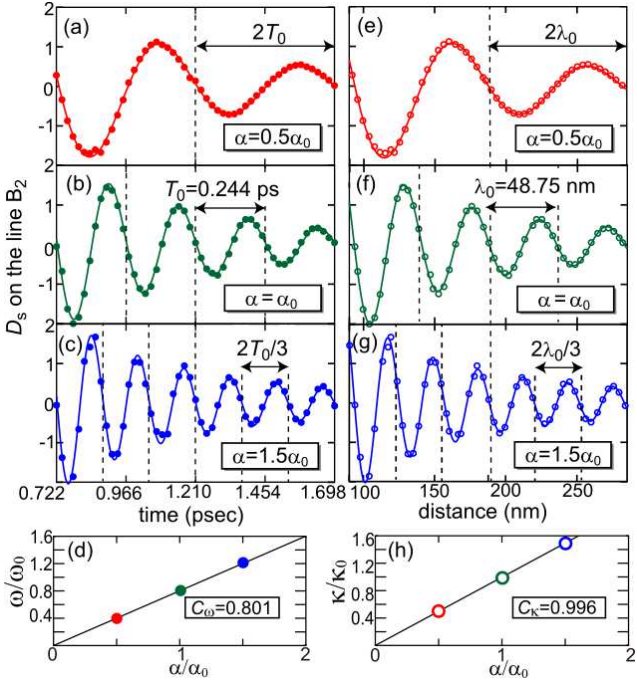}
\caption{(a)-(c) Time profiles of $D_{\rm s}$ on the line B$_2$ [see Fig.~\ref{Fig2}(e)] for various Rashba parameters, (a) $\alpha=0.5\alpha_0$, (b) $\alpha=\alpha_0$, and (c) $\alpha=1.5\alpha_0$, where the dots and the solid lines represent the simulated data and fitting curves (see text), respectively. (d) Rashba-parameter dependence of the spin-precession frequency $\omega$, which is proportional to the slope of $C_\omega$=0.801. (e)-(g) Spatial profiles of $D_{\rm s}$ on the line B$_2$ for various Rashba parameters where the horizontal axis represents the distance from a center of the slit. (h) Rashba-parameter dependence of the wave number $\kappa$ of the dynamical spin density $D_{\rm s}$, which is again proportional to the slope of $C_\kappa$=0.996.}
\label{Fig4}
\end{figure}
To investigate the properties and the Rashba-parameter dependence of the spatiotemporal oscillation of $D_{\rm s}$, the simulated quantities of $D_{\rm s}(\bm r=\bm r_{\rm max},t)$ are plotted in Figs.~\ref{Fig4}(a)-\ref{Fig4}(c) as functions of time $t$ after passing through the slit and in Figs.~\ref{Fig4}(e)-\ref{Fig4}(f) as functions of distance from the slit for various values of the Rashba parameter $\alpha$. Here $\bm r_{\rm max}$ is the position at which $D_{\rm c}$ is a maximum on the line B$_2$ at time $t$. 

Before discussing these results, let us overview the cases without the slit. When both the RSOI and the slit are absent, the charge density $D_{\rm c}$ for the Gaussian wave packet is given in the form,
\begin{eqnarray}
D_{\rm c}^{\rm free}(\bm r,t)=&&\frac{1}{2\pi\delta^2\left(1+\left(\frac{\hbar t}{2m\delta^2}\right)^2\right)} \notag \\
&&\exp\left[-\frac{(x-x_0-v_{\rm g}t)^2+y^2}{2\delta^2\left(1+\left(\frac{\hbar t}{2m\delta^2}\right)^2\right)}\right], 
\label{eq:Dcfree}
\end{eqnarray}
where $v_{\rm g}=\hbar k/m$ is the group velocity, and $k$ is the electron wave number. When the RSOI sets in, the induced effective magnetic field $\bm H_{\rm RSOI}$ gives rise to the precession of electron spins, which is described by the Heisenberg equation,
\begin{eqnarray}
\frac{d}{dt}\hat{\bm{\sigma}}(t)
=\frac{2}{\hbar}(\bm k \times\bm{\alpha})\times\hat{\bm{\sigma}}(t).
\end{eqnarray}
Solving this equation, we obtain the expression of the angular frequency $\omega$ for the spin precession as $\omega=2\alpha k/\hbar$.
This RSOI-dependent angular frequency has indeed been observed experimentally in the Zitterbewegung dynamics of electrons in the RSOI system, that is, the translational motion of electrons is accompanied by tiny oscillations in space with this specific frequency because of the feedback effect from the RSOI-induced spin precession on the real-space charge dynamics~\cite{Shen05,Schliemann06}.

The angular frequency $\omega_0$ for $\alpha$=$\alpha_0$ and $k_{\rm in}$=$k_0$ in the absence of slit is given by $\omega_0=2\alpha_0 k_0/\hbar$ because the electron wave number $k$ is always identical to the incident wave number $k_{\rm in}$(=$k_0$) when the slit is absent. On the other hand, the wave number $\kappa_0$ of the moving spin density in this case is given by $\kappa_0=\omega_0/v_{\rm g}=2m\alpha_0/\hbar^2$
where $v_{\rm g}=\hbar k_0/m$.

The data plotted in Fig.~\ref{Fig4} are obtained in the presence of both the RSOI and the slit. The time profiles in Figs.~\ref{Fig4}(a)-\ref{Fig4}(c) and the space profiles in Figs.~\ref{Fig4}(d)-\ref{Fig4}(f) are fit by using,
\begin{eqnarray}
& &D_{\rm s}(\bm{r}_{{\rm max}}(t),t)=
\frac{D_t}{1+(\sigma_t t)^2}\sin(\omega t), 
\label{eq:Dstimefit}
\\
& &D_{\rm s}(\bm{r}_{{\rm max}},t(\bm{r}_{{\rm max}}))=
\frac{D_r}{1+(\sigma_r r)^2}\sin(\kappa r),
\label{eq:Dspositionfit}
\end{eqnarray}
where $\omega$ and $\kappa$ are the angular frequency of the spin precession and the wave number of the moving spin density $D_{\rm s}$ in the presence of the slit. Here $D_t$, $\sigma_t$, $D_r$, and $\sigma_r$ as well as $\omega$ and $\kappa$ are fitting parameters. These approximate expressions are deduced by analogy to the expression for $D_{\rm c}$ in free space given by Eq.~(\ref{eq:Dcfree}), which turn out to provide an excellent fitting to the simulated data. The sine factors in Eqs.~(\ref{eq:Dstimefit}) and (\ref{eq:Dspositionfit}) are added to describe the oscillatory behavior of $D_{\rm s}$ in space and time.

The values of $\omega/\omega_0$ and $\kappa/\kappa_0$ evaluated by the fittings are plotted in Figs.~\ref{Fig4}(d) and \ref{Fig4}(h), respectively, as functions of the Rashba parameter $\alpha/\alpha_0$. In Figs.~\ref{Fig4}(d) and \ref{Fig4}(h), we find apparent proportional relations for both cases as,
\begin{eqnarray}
\frac{\omega}{\omega_0}=C_{\omega}\frac{\alpha}{\alpha_0}, 
\quad\quad
\frac{\kappa}{\kappa_0}=C_{\kappa}\frac{\alpha}{\alpha_0}, 
\end{eqnarray}
where the proportionality factors are $C_\omega$=0.801 and $C_\kappa$=0.996~($\approx$1), respectively. Consequently, we obtain the relations,
\begin{eqnarray}
\omega=C_{\omega} \frac{2\alpha k_0}{\hbar},
\label{eq:omega2}
\quad\quad
\kappa=C_{\kappa} \frac{2m\alpha}{\hbar^2} \sim \frac{2m\alpha}{\hbar^2}.
\label{eq:kappa2}
\end{eqnarray}
in the presence of both the RSOI and a slit. The wave number of incident electrons $k_{\rm in}(=k_0)$ is reduced to $k_{\rm out}=C_\omega k_{\rm in}(=C_\omega k_0)$ by a factor of $C_\omega$ after passing through the slit. Moreover, the wave number of the spin precession $\kappa$ is independent of the incident wave number $k_{\rm in}$ but is only proportional to the Rashba parameter $\alpha$. 

\begin{figure}[tb]
\includegraphics[scale=0.5]{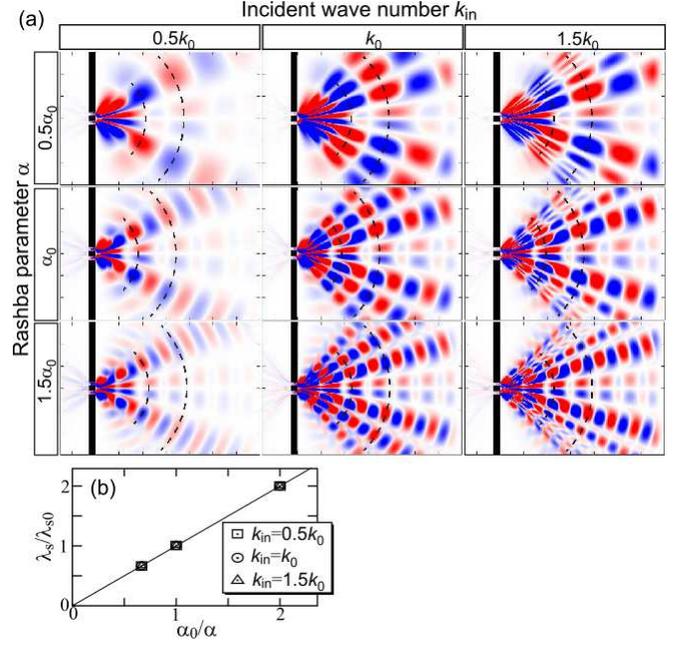}
\caption{(a) Superimposed images of overlaid snapshots of $D_{\rm s}$ after passing through the double slit obtained in the simulations for various combinations of the wave numbers $k_{\rm in}$ of the incident electron and the Rashba parameters $\alpha$. A couple of dashed lines in each panel are guides for eyes, which are drawn with the same spatial intervals. The areas shown in these figures are identical to those in Figs.~\ref{Fig2}(b) and \ref{Fig2}(e). (b) Rashba-parameter dependence of $\lambda_{\rm s}$ for various values of the incident wave number $k_{\rm in}$, where $\lambda_{\rm s0}$ is the spatial period of spin density when $\alpha=\alpha_0$ and $k_{\rm in}=k_0$.}
\label{Fig5}
\end{figure}
We now derive the spatial period $\lambda_{\rm s}$ of the oscillation in $D_{\rm s}$,
\begin{eqnarray}
\lambda_{\rm s}=\frac{2\pi}{\omega}\cdot\frac{\hbar k_{\rm out}}{m}=\frac{\pi\hbar^2}{m\alpha},
\label{eq:lambda}
\end{eqnarray}
which is inversely proportional to $\alpha$ but is independent of $k_{\rm in}$. To confirm the validity of this prediction, we performed simulations by varying the values of $\alpha$ and $k_{\rm in}$. Figure~\ref{Fig5}(a) shows superimposed images of $D_{\rm s}$ after passing through the double slit for various values of $\alpha$ and $k_{\rm in}$. Here the dashed lines indicate equidistance positions from the slit. We realize that upon varying $\alpha$, the spatial period $\lambda_{\rm s}$ is inversely proportional to $\alpha$, whereas the directions in which the enhanced oscillations of $D_{\rm s}$ occur do not change. Conversely, the spatial period $\lambda_{\rm s}$ never changes upon varying $k_{\rm in}$, whereas the directions along which the fringes appear and the number of fringes vary. These aspects can be clearly seen in Fig.~\ref{Fig5}(b) which shows plots of $\lambda_{\rm s}$ as functions of $1/\alpha$ for various values of $k_{\rm in}$. We find that all the plots are linear in $1/\alpha$ and are perfectly overlapped, which supports the predicted $1/\alpha$-linear and $k_{\rm in}$-independent behaviors of $\lambda_{\rm s}$. These results are consistent with Eq.~(\ref{eq:lambda}). The analysis gives a precise factor of proportionality as $\lambda_{\rm s}=47.2\times(\alpha/\alpha_0)^{-1}$ nm.

To summarize, we theoretically investigated the charge and spin dynamics of electrons in single- and double-slit experiments for the two-dimensional Rashba electron system. When an incident electron diffracts from a slit, its spin, which is originally polarized perpendicular to the incident direction, starts to precess around the Rashba-mediated effective magnetic field $\bm H_{\rm RSOI}$. Consequently, the spin density $D_{\rm s}$ and the charge density $D_{\rm c}$ exhibit interference fringes. Since the precession of electron spins and the spin orientations of the fringes can be controlled not only by tuning the slit parameters but also by modulating the RSOI via application of an electric gate voltage, this phenomenon may be exploited to transfer the spin information or to generate the spin-current in spintronics devices. To realize the proposed experiments, semiconductor nanofabrication techniques are required because the confinement length of the slit must be shorter than the positional uncertainty of incident electrons to observe clear interference fringes. In fact, the single-slit system was recently demonstrated experimentally in a two-dimensional electron system by using a quantum point contact~\cite{Kolashinski17}. Conversely, the double-slit experiment may require more precise fabrication, which is a challenge for future research.

This work was partly supported by JSPS KAKENHI (Grant No. 17H02924, No. 16H06345 and No. 19H00864) and Waseda University Grant for Special Research Projects (Project No. 2019C-253).


\end{document}